\documentclass[aps,prc,twocolumn,showpacs,showkeys,amsmath,amssymb,nofootinbib]{revtex4}
\usepackage{graphicx}
\usepackage{dcolumn}
\usepackage{amsmath}
\usepackage{bm}
\usepackage{float}
\usepackage{hyperref}
\usepackage{subfigure}
\usepackage{multirow}
\usepackage{subfigure}
\usepackage{rotating}
\usepackage{color}

\newcolumntype{C}[1]{>{\centering\arraybackslash}p{#1}}

\def\Tr{{\rm Tr}}
\def\arctanh{{\rm arctanh}}

\newcommand{\degr}{{}^{\tiny o}}

 \allowdisplaybreaks

\newcounter{rpl}

\newcounter{rpla}

\newenvironment{table**}[1][]{\begin{widetext}
\begin{table}[#1]
\begin{minipage}{\textwidth}
}{\end{minipage}
\end{table}
\end{widetext} }

\begin{document}

\author{Zhan-Wei Liu}
\affiliation{CSSM, School of Chemistry and Physics, University of Adelaide, Adelaide, South Australia 5005, Australia}
\author{M. E. Carrillo-Serrano}
\affiliation{CSSM and ARC Centre of Excellence for Particle Physics at the Tera-scale,\\
School of Chemistry and Physics, University of Adelaide, Adelaide SA 5005, Australia}
\author{A. W. Thomas}
\affiliation{CSSM and ARC Centre of Excellence for Particle Physics at the Tera-scale,\\
School of Chemistry and Physics, University of Adelaide, Adelaide SA 5005, Australia}

\title{Study of Kaon Decay to Two Pions}

%
%

\begin{abstract}
The weak decay of the kaon to two pions is studied within the 
model of Nambu and Jona-Lasinio (NJL Model). 
Using the standard effective weak Hamiltonian,  
both the decay amplitude arising from an intermediate state 
$\sigma$ meson and the direct decay amplitude are calculated. 
The effect of final state interactions is also included.
When the matching scale is chosen such that the decay amplitude 
with isospin $I=2$ is close to its experimental value,
our model including the $\sigma$ meson contributes up to 80\% of the total $I=0$ 
amplitude.
This supports recent suggestions that the $\sigma$ meson 
should play a vital role in explaining the $\Delta I= 1/2$ rule in 
this system. 
\end{abstract}
\pacs{13.25.Es, 13.25.-k, 12.39.-x, 12.40.-y}
\keywords{kaon decay, intermediate sigma meson state, final state interaction, NJL model}

\maketitle

\section{Introduction}\label{secInt}
The $\Delta I=1/2$ rule \cite{Gell-Mann1955,Gell-Mann1957}, notably in the $K \rightarrow \pi \pi$ decay, 
is one of the major outstanding challenges to our 
understanding of the hadronic weak interaction. It has therefore 
been studied with many different 
theoretical methods~\cite{Boyle2013,Bertolini1998,
Kambor1990,Kambor1991,Kambor:1992he,
Bijnens2009,Buras2014,Hambye1999,Hambye2003,
Cirigliano2004,Gerard2001,Giusti2004,Hambye2006,Buras2014a,
Truong1988,Crewther1986,Shifman1977,Carrasco2013,Buras2014b}.
In recent years these efforts have been extended to include 
lattice QCD studies, 
with recent results reported in Ref.~\cite{Boyle2013} and 
Refs.~\cite{Blum2012,Blum2012a}, the latter focussing on decays into 
the isospin $I = 2$ channel.

Amongst many quark model studies devoted to this problem, 
we note that in Ref.~\cite{Bertolini1998} 
the authors calculated the matrix elements up to $O(p^4)$ 
within the framework of the chiral quark model.
Using chiral perturbation theory, 
Kambor {\it et al.}~\cite{Kambor1990,Kambor1991,Kambor:1992he} 
studied the kaon decays to one loop order within SU(3).
Again, within SU(3) chiral perturbation theory, 
the effect of isospin breaking was included and one-loop results 
reported in Ref.~\cite{Cirigliano2004}.
Bijnens {\it et al.}~\cite{Bijnens2009} studied the kaon decays 
to one loop order within SU(2) chiral perturbation theory. 
NLO contributions were considered within the large $N_c$ approach 
in Refs.~\cite{Buras2014,Hambye1999,Hambye2003}.
The potentially important role of the trace anomaly in weak $K$-decays,
especially in regard to the $\Delta I=1/2$ rule, was 
discussed in Ref.~\cite{Gerard2001}.

The possible role of the charm quark in generating the observed enhancement 
was discussed in Ref.~\cite{Giusti2004}, with the authors presenting  
there the first results from lattice simulations in the SU(4) flavor limit.
In Ref.~\cite{Hambye2006} the authors studied the problem within 
the framework of a dual 5-dimensional holographic QCD model.
The possible effect of ``new physics'', specifically the effect of 
introducing a heavy colorless $Z'$ gauge boson, was discussed by 
Buras {\it et al.}~\cite{Buras2014a}.

In a recent report~\cite{Buras2014b}, Buras summarized a study of 
this rule based on the dual representation of QCD 
using the large $N_c$ expansion.
The Wilson coefficients and hadronic matrix elements were evaluated 
at different energy scales, $\mu$, 
in the early large $N_c$ studies, 
and thus the calculated value of $A_0$ was only about 10\% of 
the experimental one.
By evaluating the Wilson coefficients and hadronic matrix elements at 
the same energy scale,
the discrepancy was decreased by about 40\%. 
Moreover, the introduction of QCD penguin operators further decreased
the initial discrepancy.

The effect of final state interactions (FSI) was studied in various ways in
Refs.~\cite{Belkov1989,Locher1997,Pallante2000,Pallante2000a,Buechler2001,Isgur1990,Brown1990}.
{}For example, in Ref.~\cite{Belkov1989} the authors directly calculated 
the relevant Feynman diagrams for the meson rescattering corrections 
in chiral perturbation theory.
The Omn{\`e}s approach, which is based on dispersion relations, 
was used in Refs.~\cite{Locher1997,Pallante2000,Pallante2000a,Buechler2001}, 
while in Refs.~\cite{Isgur1990,Brown1990} the effect of FSI was
evaluated within potential models.

Of particular interest to us is the recent work by 
Crewther and Tunstall~\cite{Crewther2013,Crewther2012,Crewther2013a},  
which examined the proposal that the $\Delta I=1/2$ rule might be resolved 
if QCD were to have an infrared fixed point. This suggested that the $\sigma$ 
meson would play an especially important role. 
While the existence of the $\sigma$ meson 
has been controversial for decades, there is now convincing evidence 
of a pole in the $\pi-\pi$ scattering amplitude  
with a mass similar to that of the kaon, albeit with 
a very large width. 
Given that there is a known scalar resonance nearly degenerate with the kaon, 
it is clear that such a state 
may well play a significant role in the $K \, \rightarrow \, 2 \, \pi$ decay. 
With this motivation, we use the NJL model, together with the familiar 
operator product formulation of the non-leptonic weak interaction,  
to make an explicit calculation of the role of the $\sigma$ meson in 
the decay $K \, \rightarrow \, 2 \, \pi$, with the aim of 
clarifying its role in the $\Delta I \, = \, 1/2$ rule.
Section \ref{secFrm} gives details of the calculation of the $\sigma$ 
contribution, while the direct decay to pions is found in sect.~\ref{secDir}.
The numerical results and discussion are given in sect.~\ref{secNRD}.


\section{Calculation of Kaon Decay including the $\sigma$ meson}\label{secFrm}
{}Following the standard conventions we label the $K$ decay to two 
pions with isospin zero as $A_0$ and with isospin two as $A_2$ \cite{Blum2012},
\begin{equation}
  A_I \equiv \frac{1}{\sqrt2} \langle (\pi\pi)_I|K^0\rangle,
  \qquad I=0, 2.
  \label{eqAI}
\end{equation}
As explained earlier, for the former we calculate the contribution 
from two different mechanisms; first, the weak transition from $K$ 
to a $\sigma$ meson followed by the decay of the $\sigma$ to two pions
and second, the direct decay to two pions. For 
$A_2$ only the latter path is available.

\begin{figure}[!htbp]
 \centering
 \scalebox{0.8}{\includegraphics{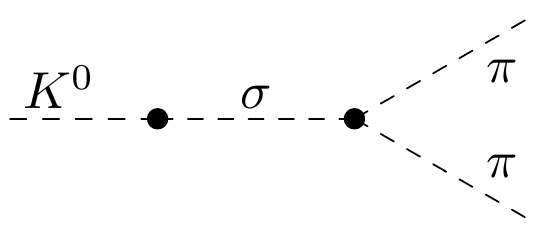} }
 \caption{Contribution of $\sigma$ meson to $K\to\pi\pi$.}
 \label{figKPi}
 \end{figure}

In the absence of final state interactions (which will be included later),
the first contribution to $A_0$, as illustrated  
in Fig.~\ref{figKPi}
is written:
\begin{equation}
A_0^{\sigma,0}=-\sqrt{\frac{3}{2}}g_{K\sigma}\times 
\Delta_\sigma\times \gamma~(\frac{m_K^2}{2}-m_\pi^2) \, ,
\label{eqA0}
\end{equation}
where $g_{K\sigma}$ is the coupling for the $K\sigma$ transition, 
$\Delta_\sigma$ is the propagator of the $\sigma$ meson  
and $\gamma$ is the $\sigma\pi\pi$ 
coupling~\cite{Crewther2012,Harada1996}
\begin{eqnarray}
\mathcal L_{K\sigma}&=&g_{K\sigma}K^0_S \sigma
= \frac{g_{K\sigma}}{\sqrt2}\bar K^0 \sigma  + 
\frac{g_{K\sigma}}{\sqrt2}K^0 \sigma,
\\
\mathcal L_{\sigma\pi\pi}&=&
-\frac{\gamma}{\sqrt2}\sigma \partial_\mu \vec\pi\cdot \partial^\mu \vec\pi
\label{eqLHd}
\end{eqnarray}
and we have neglected the effect of CP-violation.

We employ the NJL model with 
dimensional regularization to describe the 
structure of these mesons. The coupling of the $\sigma$ to the pions 
is also determined  within the NJL model. 
Finally, the effective 
Hamiltonian describing the non-leptonic weak interaction is obtained 
using the standard operator product expansion. We now briefly summarise 
each of these parts of the calculation.


\subsection{NJL model\label{Sec:NJL}}
Our work uses the NJL formalism based upon SU(3)-flavour symmetry.
After Fierz transformation,
the Lagrangian density can be written in the meson channels. In this form 
the contributions from the different types of meson can be read directly
~\cite{Klevansky1992,Vogl1991}.
This has recently been used in the computation
of the kaon and pion form factors~\cite{Ninomiya:2014kja}, 
as well as the study 
of SU(3)-flavour symmetry in the baryon octet~\cite{Carrillo-Serrano:2014zta}.
Those studies included the breaking of SU(3) chiral symmetry with the use of
different masses for the constituent light quarks (up and down) 
and the constituent
strange quark.

Here we include different couplings for the scalar ($\sigma$) and pseudoscalar
mesons (pion and kaon), modifying the NJL Lagrangian density as follows:
\begin{align}
\tilde{\mathcal{L}}_{I}^{NJL} &= G_\sigma\left[\tfrac{2}{3}\left(\bar{\psi}\psi\right)^{2} 
+ \left(\bar{\psi}\,\boldsymbol{\lambda}\,\psi\right)^2\right]\nonumber\\
& - G_{\pi}\left[\tfrac{2}{3}\left(\bar{\psi}\,\gamma_{5}\,\psi\right)^2 
+ \left(\bar{\psi}\,\gamma_{5}\,\boldsymbol{\lambda}\,\psi\right)^2\right] \, ,
\label{Eq:Lag}
\end{align}
where the eight Gell-Mann SU(3)-flavor matrices are 
represented as $\boldsymbol{\lambda}$. 
This modified NJL lagrangian density 
preserves ${\rm SU_V(3)\otimes U_V(1)}$ symmetry. 

Since NJL is an effective model, it needs to be regularized. 
We chose dimensional regularization for consistency with the computation
of the Wilson coefficients when the electroweak interaction
is included (Sec.~\ref{Sec:EffWeakHam}). The value of the 
energy scale $\mu$ is constrained by requiring stability of the 
Wilson coefficients (Fig.~\ref{figWilson}).
With the Lagrangian density of Eq.~\eqref{Eq:Lag} the Gap equation for
the constituent light quark $M_l$ comes from the scalar interaction
term:
\begin{equation}
M_l=m_l+48iG_{\sigma}\int\frac{d^4k}{(2\pi)^4}\frac{1}{k^2-M_l^2+i\epsilon}.
\label{Eq:Gap}
\end{equation}
where $m_l$ is the mass of the current light quark.

\begin{figure}[t]
 \includegraphics[width=\columnwidth]{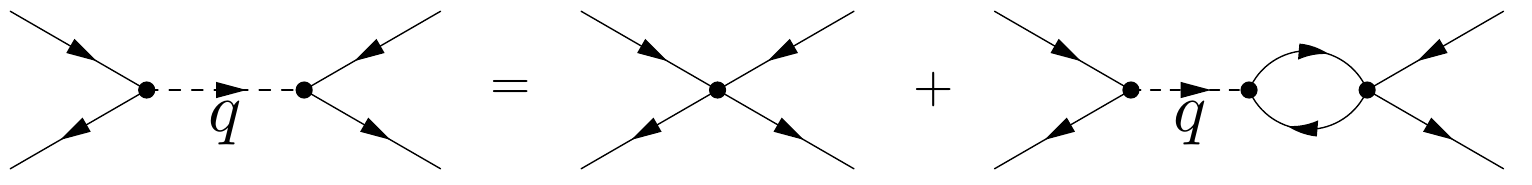} 
 \caption{Diagrammatic representation of the inhomegeneous 
Bethe-Salpeter Equation for the 
different quark-antiquark bound states (mesons) of total 4-momentum $q$.}
 \label{BSEqn}
 \end{figure}

With $\tilde{\mathcal{L}}_{I}^{NJL}$ we follow the standard method 
of solving the Bethe-Salpeter 
equations (BSE) for the quark antiquark bound 
states (mesons)~\cite{Klevansky1992,Vogl1991}. 
The diagram describing this BSE in the NJL model is shown 
in Fig.~\ref{BSEqn}, and its solutions 
are given by the following reduced t-matrices:
\begin{equation}
\mathcal{T}_{j}\left(q\right) 
= \frac{- 2iG_{j}}{1\pm2G_{j}\Pi_{j}\left(q^{2}\right)}.
\label{Eq:TMatrix}
\end{equation}
Here, the polarization, $\Pi_{j}\left(q^{2}\right)$, represents 
the quark-antiquark loops that 
appear in the diagram for the BSE ($j = \sigma$-meson, pion or kaon).
with the + and - signs corresponding to the 
pion and $\sigma$ respectively.
Their analytic expressions are
\begin{equation}
\Pi_{\sigma}(q) = 6i\int \frac{d^{4}k}{\left(2\pi\right)^{4}} \Tr\left[S_{q_{1}}\left(k\right)S_{q_{2}}\left(k+q\right)\right],
\end{equation}
and
\begin{equation}
\Pi_{\pi(K)}(q) = 6i\int \frac{d^{4}k}{\left(2\pi\right)^{4}} 
\Tr\left[\gamma_{5}S_{q_{1}}\left(k\right)\gamma_{5}S_{q_{2}}
\left(k+q\right)\right] \, ,
\end{equation}
where $\Tr$ is a trace in Lorentz indices (the traces over color 
and flavour having already been taken) and $S_{i}$ are 
the constituent quark propagators.
For the $\sigma$ and pion the two propagators contain 
the same light quark masses, whereas for the kaon 
case their masses are different.
The explicit expressions for $\Pi_j(q^2)$ in dimensional 
regularization are

\begin{align}
\Pi_{\pi(K)}(q) &= -12\left[J_0^i(M_1) + J_0^1(M_2)\right.\nonumber\\
 &\left.- (q^2-(M_2 - M_1)^2)J_0^F(M_1,M_2,q^2)\right], 
\end{align}
and
\begin{equation}
\Pi_{\sigma}(q) = 24J_0^i(M_l) - 12(q^2-4M_l^2)J_0^F(M_l,M_l,q^2), 
\end{equation}
where $M_1 = M_2 = M_l$ for the pion, and $M_1 = M_l$ and $M_2 = M_s$ for the kaon. The integrals 
$J_0^i(M)$ and $J_0^F(M_1,M_2,q^2)$ are given in Appendix~\ref{secQ56}.  

The pole position of $\mathcal{T}_{j}\left(q\right)$ corresponds 
to the mass of each of 
the mesons, $(j)$, which is evident if one examines 
the expression for $\mathcal{T}_{j}\left(q\right)$ 
in pole approximation~\cite{Klevansky1992}
\begin{equation}
\mathcal{T}_{j}\left(q\right) \to \frac{- i g_{j}^{2}}{q^{2}-m_{j}^{2}},
\end{equation}
where $g_{j}$ is the effective quark-meson coupling, given by
\begin{equation}
g_{j}^{2} = \mp\left. \left(\frac{\partial \Pi_{j}}{\partial q^{2}}
\right)^{-1}\right|_{q^{2} = m_{j}^{2}} \, .
\label{Eq:EffCoup}
\end{equation}
The - and + signs correspond to the pion (kaon) and the $\sigma$, respectively, 
with the sign difference coming from Eq.~\ref{Eq:TMatrix}.

Here we assume degenerate masses for the constituent light 
quarks ($M_{l} = M_{u} = M_{d}$). 
The mass of the $\sigma$-meson ($m_{\sigma}$) is taken 
to lie in the range 520 - 600 MeV. With
the gap equation (Eq.~\ref{Eq:Gap}), including a current light 
quark mass $m_l$ of 5 Mev,
and the equation for the mass of the $\sigma$-meson 
(pole position in Eq.~\ref{Eq:TMatrix}), 
we fit $G_{\sigma}$ and $M_l$. Our result for $M_l$ is in 
reasonable agreement with
Ref.~\cite{Hatsuda:1985ey}, where it was shown that $m_{\sigma}\approx 2M_{l}$.
$G_\pi$ is chosen to reproduce the physical $m_\pi$, and $M_s$ to reproduce
the kaon mass $m_K$. Finally the effective couplings, $g_j^2$, 
are computed with
Eq.~\ref{Eq:EffCoup}. The results for $m_{\sigma} = 520$, $560$ 
and $600$ MeV are summarized
in Table~\ref{Tab:1}. The negative sign of the Lagrangian 
couplings is a feature of dimensional 
regularization in the NJL model~\cite{Inagaki:2007dq}. 
We also stress that the difference between
$G_{\sigma}$ and $G_{\pi}$ is of the order of 10\%. 

The complication associated with such a model, when one 
needs to match to operators that are defined at some renormalization 
scale, is that the scale associated with a valence-dominated quark model 
is typically quite low. For example, extensive studies of parton 
distribution functions within the NJL 
model~\cite{Mineo:2003vc,Cloet:2005pp,Cloet:2006bq} (as well as other
valence-dominated quark 
models~\cite{Schreiber:1990ij,Diakonov:1996sr}) typically 
lead to a matching scale of order 0.4-0.5 GeV. This is rather low and one 
therefore needs to check the reliability of the effective weak couplings 
at such a scale. We address this below.

\begin{table}[tbp]
\caption{Model parameters: $\mu$ and all the masses are in units of GeV, the couplings $G_{\sigma}$ and $G_{\pi}$ are in units of GeV$^{-2}$, the $\sigma$ to $\pi-\pi$ coupling $\gamma$ is in units of GeV$^{-1}$, and the effective couplings $g_{i}$ are dimensionless.} 
\label{Tab:1}
\addtolength{\tabcolsep}{0.0pt}
\begin{tabular}{ccccccccc}
\hline\hline
\multicolumn{9}{c}{$m_{\sigma} = 0.520$}\\
\hline
$\mu$ & $G_{\sigma}$ & $G_{\pi}$ & $M_{l}$  & $M_s$  & $g_{\sigma}^{2}$ & $g_{K}^{2}$ & $g_{\pi}^{2}$ & $|\gamma|$\\
\hline
0.48  &  -21.35    & -23.72    &  0.261   & 0.549   & 4.629   & 16.174    &  9.975    &  3.737   \\
0.50  &  -20.60    & -22.93    &  0.261   & 0.539   & 4.502   & 13.920    &  9.394    &  3.762   \\
0.70  &  -15.93    & -18.00    &  0.261   & 0.514   & 3.671   &  7.472    &  6.347    &  3.761   \\
\hline
\multicolumn{9}{c}{$m_{\sigma} = 0.560$}\\
\hline
0.48  &  -19.766    & -21.564    &  0.281   & 0.575   & 4.852   & 20.795    &  11.370    &  4.234   \\
0.50  &  -19.016    & -20.794    &  0.281   & 0.589   & 4.713   & 16.614    &  10.621    &  4.262   \\
0.70  &  -14.483    & -16.063    &  0.281   & 0.527   & 3.811   &  8.111    &   6.885    &  4.228   \\
\hline
\multicolumn{9}{c}{$m_{\sigma} = 0.600$}\\
\hline
0.48  &  -18.475    & -19.866    &  0.301   & 0.613   & 5.081   & 30.983    &  13.048    &  4.703   \\
0.50  &  -17.726    & -19.105    &  0.301   & 0.583   & 4.929   & 20.832    &  12.072    &  4.737   \\
0.70  &  -13.285    & -14.517    &  0.301   & 0.541   & 3.952   &  8.810    &   7.466    &  4.686   \\
\hline\hline
\end{tabular}
\end{table}

\subsection{Coupling between $\sigma$ and $\pi$-$\pi$}
\begin{figure}[!htbp]
 \centering
 \includegraphics[width=0.6\columnwidth]{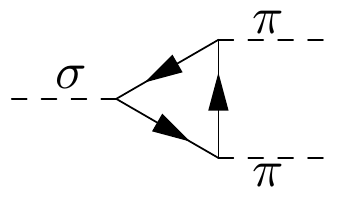} 
 \caption{Illustration of the $\sigma\to \pi\pi$ process at quark level, where the solid lines represent 
$u$ or $d$ quarks.}
 \label{figSPiPi}
 \end{figure}

We obtain the coupling, $\gamma$, between $\sigma$ and $\pi$-$\pi$ 
within the NJL model.
To that end one should calculate the amplitudes of the $\sigma\to \pi\pi$ 
process at both quark and hadron levels,
and match the results.
At quark level, the amplitude can be obtained from Fig. \ref{figSPiPi} with the masses and couplings derived within NJL model. 
At the hadron level, the amplitude can be easily given from 
the effective Lagrangian $\mathcal L_{\sigma\pi\pi}$  in Eq. (\ref{eqLHd}),
\begin{equation}
T_{\sigma\to \pi^+\pi^-}^{\rm Hadron~Level}=i \sqrt2 
\gamma p_{\pi^+} p_{\pi^-} \, .
\end{equation}

We match both amplitudes at a centre-of-mass energy of 
the system $\sqrt{s}=m_K$, 
since the coupling $\gamma$ would be used to study the decay of kaon.
The amplitude from Fig.~\ref{figSPiPi} at the quark level 
is energy-scale dependent, 
and therefore $\gamma$ also runs as the energy scale $\mu$ changes 
within our model.
However, $\gamma$ is rather insensitive to $\mu$, as we see from 
the numerical results in Table~\ref{Tab:1}.
\subsection{Effective weak Hamiltonian} \label{Sec:EffWeakHam}
Here we need the $\Delta S=1$ effective Lagrangian of the electroweak 
interaction~\cite{Buchalla1996}
\begin{equation}
\mathcal H_{\mbox{eff}}=\frac{G_F}{\sqrt2}V^*_{us}V_{ud}\sum_{i=1}^6
\left( z_i(\mu)-\frac{V_{ts}^*V_{td}}{V_{us}^*V_{ud}}y_i(\mu) \right) Q_i \, ,
\label{eqHew}
\end{equation}
where $V_{xy}$ is the relevant CKM matrix element, $G_F$ is the Fermi 
coupling constant  
and the four-quark operators, $Q_i$, are:
\begin{eqnarray}
 &&
 Q_1=\bar s_\alpha\gamma_\mu(1-\gamma_5)u_\beta~\bar u_\beta
\gamma^\mu(1-\gamma_5)d_\alpha,
 \\&&
 Q_2=\bar s_\alpha\gamma_\mu(1-\gamma_5)u_\alpha ~\bar u_\beta
\gamma^\mu(1-\gamma_5)d_\beta,
 \\&&
Q_3=\bar s_\alpha\gamma_\mu(1-\gamma_5)d_\alpha~\bar q_\beta
\gamma^\mu(1-\gamma_5)q_\beta,
 \\&&
 Q_4=\bar s_\alpha\gamma_\mu(1-\gamma_5)d_\beta ~\bar q_\beta
\gamma^\mu(1-\gamma_5)q_\alpha,
 \\&&
Q_5=\bar s_\alpha\gamma_\mu(1-\gamma_5)d_\alpha~\bar q_\beta
\gamma^\mu(1+\gamma_5)q_\beta,
 \\&&
 Q_6=\bar s_\alpha\gamma_\mu(1-\gamma_5)d_\beta ~\bar q_\beta
\gamma^\mu(1+\gamma_5)q_\alpha.
 \label{eqEOQ}
\end{eqnarray}
The Wilson coefficients, $z_i(\mu)$ and $y_i(\mu)$, have been calculated 
up to the next to leading order
using perturbative QCD~\cite{Buchalla1996}.
Since $V_{ts}^*V_{td}/V_{us}^*V_{ud}$ is relatively small,
we will only keep the contribution of the terms with $z_i$.

\begin{figure}[!htbp]
 \centering
 \scalebox{0.5}{\includegraphics{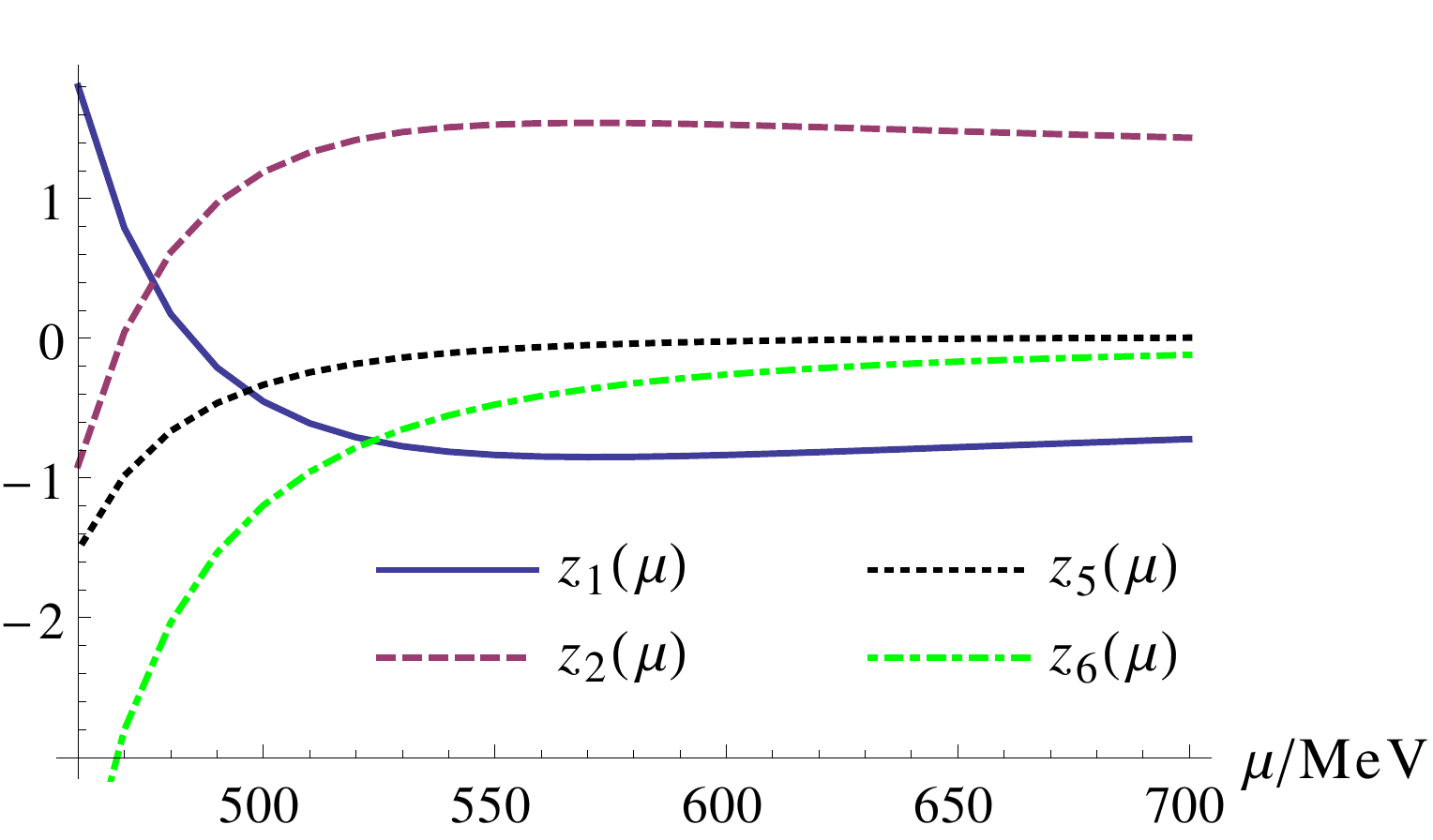} }
 \caption{Wilson Coefficients, $z_i(\mu)$. The abscissa represents 
the energy scale, $\mu$, in units of MeV.}
 \label{figWilson}
 \end{figure}

In order to investigate the potential model dependence in matching the 
renormalization group scale of the operators to the NJL model, in 
{}Fig.~\ref{figWilson} we show the variation of the coefficients 
$z_i(\mu)$ as $\mu$ varies from 700 to 450 MeV.
We can see that these Wilson coefficients vary particularly quickly 
as $\mu$ drops below 480 MeV and clearly, if one wants reliable results,  
one should not choose a scale far below this limit.
%

\subsection{Coupling for the $K \sigma$ transition}
\begin{figure}[!htbp]
 \centering
 \includegraphics[width=0.6\columnwidth]{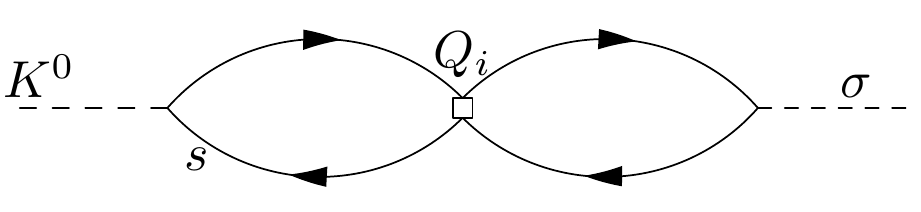} 
 \caption{Illustration of the $K\sigma$ transition, where the solid line 
with an $s$ represents $s$ quark and the other solid lines represent 
$u$ or $d$ quarks.}
 \label{figQ}
 \end{figure}

With the Wilson coefficients and the NJL model explained, we can proceed 
with the calculation of the weak $K$ to $\sigma$ transition 
amplitude, $g_{K\sigma}$
\begin{equation}
g_{K\sigma}=\sqrt2 \langle \sigma|\mathcal H_{\mbox{eff}}| K^0 \rangle \, ,
\label{eqgKsig}
\end{equation}
as illustrated in Fig.~\ref{figQ}.
Here we simply assume that the quarks appearing in the QCD 
operators $Q_i$ are the same as the NJL 
quark operators of the corresponding flavors with the energy scale $\mu$ 
lying in some region
not yet accurately specified. 
Therefore, we first show our results for $\mu$ in the 
range 0.48$\sim$0.70 GeV 
and then use the numerical results to identify the optimal region.
This is shown in Section~\ref{secNRD}.

The corresponding matrix elements are evaluated with dimensional 
regularization using 
modified minimal subtraction in order to be consistent with the 
relevant Wilson coefficients, $z_i(\mu)$.
We find that the contributions of $Q_1\sim Q_4$ to $g_{K\sigma}$ vanish,
with only $Q_5$ and $Q_6$ contributing to $g_{K\sigma}$ in our results.
In a more sophisticated model, where the masses of the constituent quarks 
and the couplings between the mesons and quark pairs
were momentum dependent,
the operators $Q_1$to $Q_4$ would also contribute to $g_{K\sigma}$.
The full expressions for the $K\sigma$ transition 
amplitude are given in Appendix \ref{secQ56}.
%


\section{Direct Decay to Pions}\label{secDir}
\begin{figure}[!htbp]
 \centering
 \subfigure[]{
 \scalebox{0.8}{\includegraphics{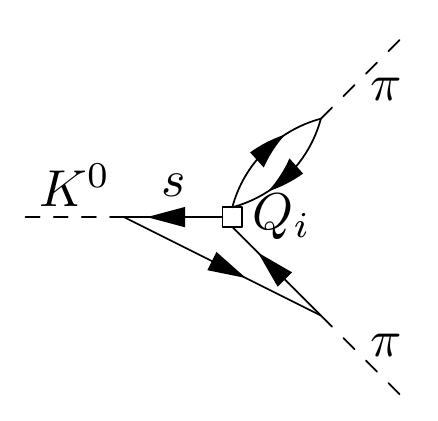} }
 \label{figKDa}
 }
 \subfigure[]{
 \scalebox{0.8}{\includegraphics{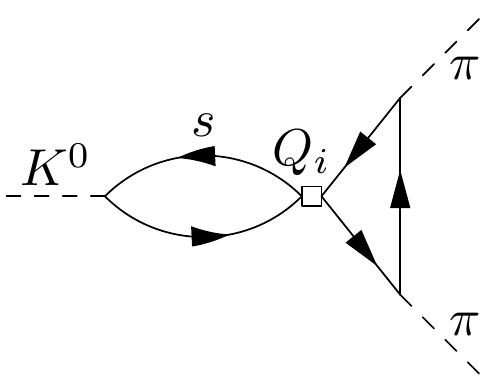} }
 \label{figKDb}
 }
  \subfigure[]{
 \scalebox{0.8}{\includegraphics{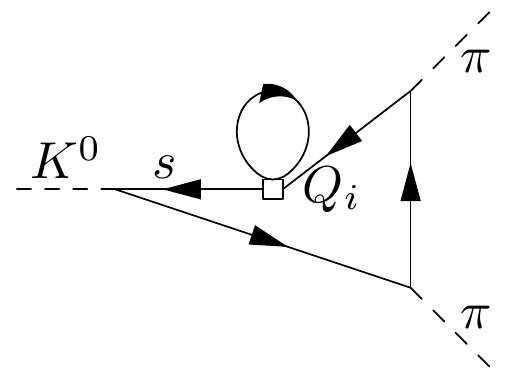} }
 \label{figKDc}
 }
 \caption{The possible diagrams for the direct weak decay. The solid line with $s$ represents $s$ quark, other solid line represents $u$ or $d$ quark. For the operator $Q_1$ and $Q_2$, only Fig. \ref{figKDa} contributes in the NJL model.}
 \label{figKpipiD}
 \end{figure}

The second mechanism contributing to the decay $K \rightarrow \pi \, \pi$ 
proceeds directly to two pions, as illustrated in Fig.~\ref{figKDa}.
Since the Wilson coefficients $z_1$ and $z_2$ are much larger than others,
we only consider the contributions of $Q_1$ and $Q_2$.
Once again the diagrams are calculated with dimensional regularization and
modified minimal subtraction.
After calculation, we find that only Fig.~\ref{figKDa} contributes 
to our results in the NJL model.

\section{Final State Interaction}
We denote the amplitudes corresponding to the diagrams shown in 
{}Figs.~\ref{figKPi} and ~\ref{figKpipiD},  
without the contribution of the final state interaction, 
as $A_I^0$($A_I^{\sigma,0}$ and $A_I^{D,0}$).
We must also consider the effect of final state 
interactions(FSI), which we treat using the method of 
Refs.~\cite{Locher1997,Pallante2000}
\begin{eqnarray}
  A_I&=&A_I^{0}\times F_I(m_K^2)\nonumber\\
  &\approx& A_I^{0}\times\exp(\frac{m_K^2}{\pi}\int_{4m_\pi^2}^\infty \frac{\delta^I(s')}{s'(s'-m_K^2)}ds')\nonumber\\
 & \approx& A_I^{0}\times\exp(\frac{m_K^2}{\pi}\int_{4m_\pi^2}^{\rm 1.0 GeV} \frac{\delta^I(s')}{s'(s'-m_K^2)}ds'),\qquad
  \label{eqAIDFSI}
\end{eqnarray}
where $\delta^I(s)$ is the phase shift for pion-pion scattering with 
isospin $I$  
and we take the values of $\delta^I(s)$ from Ref.~\cite{Dai2012}.
This yields the result:
\begin{equation}
  F_0(m_K^2)\approx1.4e^{i\,40\degr}, \quad
  F_2(m_K^2)\approx0.94e^{-i\,7.2\degr }.
  \label{eqF02}
\end{equation}
%


\section{Numerical Results and Discussion}\label{secNRD}
As we have explained, in this work $A_0$ contains two contributions,
the first, $A_0^\sigma$, involving the coupling to the $\sigma$ meson  
and the second, $A_0^D$, involving the direct decay to pions.
Since, in the NJL model, $A_0^\sigma$ involves 
the weak operators $Q_5$ and $Q_6$, 
while $A_0^D$ involves $Q_1$ and $Q_2$,
their contributions can be added with no worry about double counting
$|A_0|=|A_0^\sigma|+|A_0^D|$.

We list the $K$-$\sigma$ coupling and the decay amplitudes 
with $m_\sigma=520, 560, 600 ~{\rm MeV}$, 
as a function of the matching scale, $\mu$, 
in Tables~\ref{TBPHY520}, \ref{TBPHY560}, and \ref{TBPHY600}.
From these Tables one sees that the decay amplitudes are 
sensitive to both
$\mu$ and $m_\sigma$.
\begin{table*}[tp] 
\caption{the $K$-$\sigma$ coupling and the decay amplitudes with $m_\sigma=520~{\rm MeV}$.}\label{TBPHY520}
\begin{tabular}{c|cccccccc|cc|cc}
\hline
$\mu~(\rm{GeV})$	&	0.7	&	0.6	&	0.5	&	0.49	&	0.489	&	0.488	&	0.487	&	0.486	&	0.485	&	0.484	&	0.483	&	0.482	\\
\hline
$|g_{K\sigma}|, ~({\rm keV^2})$	&	703	&	1365	&	5184	&	6435	&	6584	&	6739	&	6898	&	7063	&	7233	&	7410	&	7592	&	7781	\\
\hline
$|A_0^\sigma|~({\rm eV})$	&	22	&	42	&	159	&	197	&	201	&	206	&	211	&	216	&	221	&	226	&	232	&	237	\\
$|A_0^D|~({\rm eV})$	&	79	&	73	&	54	&	49	&	48	&	48	&	47	&	46	&	46	&	45	&	44	&	44	\\
$|A_2^D|~({\rm eV})$	&	37	&	37	&	26	&	20	&	19	&	18	&	17	&	16	&	15	&	14	&	13	&	11	\\
\hline
$|A_0^\sigma|+|A_0^D|~({\rm eV})$	&	101	&	116	&	213	&	246	&	250	&	254	&	258	&	262	&	267	&	271	&	276	&	281	\\
$(|A_0^\sigma|+|A_0^D|)/|A_2^D|$	&	2.7	&	3.1	&	8.1	&	13	&	13	&	14	&	15	&	16	&	18	&	20	&	22	&	25	\\
\hline
\end{tabular}
\end{table*}
\begin{table*}[tp] 
\caption{the $K$-$\sigma$ coupling and the decay amplitudes with $m_\sigma=560~{\rm MeV}$.}\label{TBPHY560}
\begin{tabular}{c|ccccccc|cc|ccc}
\hline
$\mu~(\rm{GeV})$	&	0.7	&	0.6	&	0.5	&	0.49	&	0.489	&	0.488	&	0.487	&	0.486	&	0.485	&	0.484	&	0.483	&	0.482	\\
\hline
$|g_{K\sigma}|, ~({\rm keV^2})$	&	1063	&	2031	&	7423	&	9096	&	9290	&	9489	&	9694	&	9903	&	10118	&	10337	&	10561	&	10789	\\
\hline
$|A_0^\sigma|~({\rm eV})$	&	13	&	24	&	88	&	108	&	110	&	112	&	115	&	117	&	120	&	122	&	125	&	127	\\
$|A_0^D|~({\rm eV})$	&	82	&	74	&	54	&	50	&	50	&	49	&	49	&	49	&	48	&	48	&	47	&	47	\\
$|A_2^D|~({\rm eV})$	&	39	&	40	&	27	&	19	&	18	&	17	&	16	&	15	&	13	&	12	&	11	&	9.2	\\
\hline
$|A_0^\sigma|+|A_0^D|~({\rm eV})$	&	94	&	99	&	143	&	158	&	160	&	162	&	164	&	166	&	168	&	170	&	172	&	174	\\
$(|A_0^\sigma|+|A_0^D|)/|A_2^D|$	&	2.4	&	2.5	&	5.3	&	8.2	&	8.8	&	9.5	&	10	&	11	&	12	&	14	&	16	&	19	\\
\hline
\end{tabular}
\end{table*}
\begin{table*}[tp] 
\caption{the $K$-$\sigma$ coupling and the decay amplitudes with $m_\sigma=600~{\rm MeV}$.}\label{TBPHY600}
\begin{tabular}{c|ccccc|cc|ccccc}
\hline
$\mu~(\rm{GeV})$	&	0.7	&	0.6	&	0.5	&	0.49	&	0.489	&	0.488	&	0.487	&	0.486	&	0.485	&	0.484	&	0.483	&	0.482	\\
\hline
$|g_{K\sigma}|, ~({\rm keV^2})$	&	1457	&	2728	&	9312	&	10818	&	10902	&	10922	&	10736	&	10535	&	10783	&	11041	&	11310	&	11589	\\
\hline
$|A_0^\sigma|~({\rm eV})$	&	11	&	21	&	72	&	83	&	84	&	84	&	83	&	81	&	83	&	85	&	87	&	89	\\
$|A_0^D|~({\rm eV})$	&	83	&	75	&	54	&	52	&	52	&	52	&	52	&	52	&	52	&	52	&	52	&	52	\\
$|A_2^D|~({\rm eV})$	&	41	&	42	&	28	&	18	&	17	&	16	&	14	&	12	&	11	&	9.1	&	7.3	&	5.5	\\
\hline
$|A_0^\sigma|+|A_0^D|~({\rm eV})$	&	94	&	96	&	126	&	135	&	136	&	136	&	135	&	133	&	135	&	137	&	139	&	141	\\
$(|A_0^\sigma|+|A_0^D|)/|A_2^D|$	&	2.3	&	2.3	&	4.5	&	7.4	&	8	&	8.7	&	9.6	&	11	&	13	&	15	&	19	&	26	\\
\hline
\end{tabular}
\end{table*}

In Refs.~\cite{Buras2014,Buras2014b}, 
the authors used the $\overline{\rm MOM}$ scheme to evolve 
the Wilson coefficients and hadronic matrix elements to 
the same energy scale.
In order to match the energy scales,
the Wilson coefficients were evolved from $\mu=O(M_W)$ 
to $\mu=O(0.6\sim 1~{\rm GeV})$ in the quark-gluon picture,
while the hadronic matrix elements were evolved from $\mu=O(m_\pi)$ 
to the same scale $\mu=O(0.6\sim 1~{\rm GeV})$ in the meson picture.
$|A_0|/|A_2|$ was found to lie in the range 12.5$\sim$14.9 as $\mu$ 
varied from 0.6$\sim$1 GeV, if only the contributions from 
$Q_1$ and $Q_2$ were included. 

We notice that the $\mu$ dependence of their results was  
smaller than what we have found.
Here both the hadronic matrix elements and the 
Wilson coefficients are evaluated 
with dimensional regularization and modified minimal subtraction. 
(As an extension of the present work it would be interesting 
to attempt to further reduce the $\mu$-dependence
by including higher order loop corrections.)
Within the present work, as in many other applications of valence dominated 
quark models, the model is assumed to represent QCD at a scale 
at which the gluons are effectively frozen out as degrees of freedom 
and valence quarks interacting through a chiral effective Lagrangian 
dominate the dynamics. Thus the best one can do is to match the scale 
of the effective weak Hamiltonian to the scale at which the NJL model 
best matches experiment, which seems to be around $0.4-0.5$ GeV.

We note that, in addition to the processes included here, there 
are also diagrams which are disconnected if the gluon lines are removed
(usually just called disconnected diagrams for short).
While such disconnected diagrams can contribute to $A_0$,
they do not naturally appear within the NJL model and we omit them here.
Since $A_2^D$ is not contributed by the disconnected diagrams,
we use it to fix the energy scale $\mu$.

As we already noted earlier, in order that the evolution of the 
Wilson coefficients is under control, the matching scale, $\mu$,  
should not be lower than about 480 MeV. This creates some tension as the 
scale associated with the NJL model, when matching to phenomenological 
parton distribution functions, tends to be nearer 400-450 MeV.
Fortunately, we see from Tables~\ref{TBPHY520}, \ref{TBPHY560}  
and \ref{TBPHY600} that if
we choose $\mu$ to be in the range 0.484$\sim$0.488 GeV,
$A_2^D$ (which does not involve the $\sigma$ meson) actually lies 
very close to its experimental value, 14.8 eV. 
We allow a small variation of $\mu$ for different values of 
$m_\sigma$ in order to calculate $A_0$.
For $m_\sigma=520, 560, 600~{\rm MeV}$, we choose $\mu$ to be in the 
range 0.484$\sim$0.485 GeV,
0.485$\sim$0.486 GeV, and 0.487$\sim$0.488 GeV, respectively.

With $\mu$ fixed in the range where the empirical value of $A_2$ 
is reproduced, 
one notices that $|A_0|=|A_0^\sigma|+|A_0^D|$ lies in the 
range 135$\sim$270 eV,
as $m_\sigma$ varies over the range 520$\sim$600 MeV.
$A_0$ is close to the experimental value of 332 eV at $m_\sigma=520~{\rm MeV}$.
From Tables \ref{TBPHY520}, \ref{TBPHY560}, and \ref{TBPHY600}, 
we notice that $A_0^\sigma$ is sensitive to the choice of 
$m_\sigma$ because $A_0^\sigma\propto 1/(m_K-m_\sigma)$,
while  $A_0^D$ and $A_2^D$ are not sensitive to it. 
$A_0^\sigma$ decreases as $m_\sigma$ moves away from $m_K$.

In view of the uncertainties in matching the model scale to the scale of 
the weak effective Hamiltonian, it is unrealistic to expect to 
obtain a prediction for the decay amplitudes. 
Nevertheless, our calculation clearly confirms  
that the $\sigma$ meson does indeed play an important role in $A_0$,
since it contributes up to 65\% of the final value, while 
the direct decay process contributes a mere 15\%.

\section*{Acknowledgments}
We would like to thank R.~J.~Crewther and L.~C.~Tunstall (AWT) 
and A.~G.~Williams (ZWL) for helpful discussions. 
This work was supported by the Australian Research Council through 
the Centre of Excellence in Particle Physics at the Terascale and 
through an Australian Laureate Fellowship (AWT).

\appendix

\begin{widetext}
\section{expressions for the $K\sigma$ transition coupling}\label{secQ56}
One can obtain the $K\sigma$ transition coupling $g_{K\sigma}$ with Eq. (\ref{eqHew}), Eq. (\ref{eqgKsig}) 
and the matrix elements of $Q_i$.
The matrix elements of $Q_1$ to $Q_4$ vanish in NJL model, and those of $Q_5$ and $Q_6$ are expressed as
\begin{eqnarray}
&&  \langle \sigma|Q^5| K^0 \rangle=\frac13\langle \sigma|Q^6| K^0 \rangle\nonumber\\
&=&\sqrt2 g_{K^0\bar d s}g_{\sigma\bar q q}\times
\nonumber\\
&&\left\{\right.
48\left[(2 m_d^2-p^2) J^F_0(m_d,m_d,p^2)+2 p^2 J^F_{21}(m_d,m_d,p^2)+8 J^F_{22}(m_d,m_d,p^2)\right]
\nonumber\\
&&\quad\quad \times
\left[m_d m_s J^F_0(m_d,m_s,p^2)-p^2 J^F_{11}(m_d,m_s,p^2)-p^2 J^F_{21}(m_d,m_s,p^2)
      -4 J^F_{22}(m_d,m_s,p^2)\right]\nonumber\\
&&
+\frac{3}{2\pi^2}(6m_d^2-p^2)
\left[m_d m_s J^F_0(m_d,m_s,p^2)-p^2J^F_{11}(m_d,m_s,p^2)-p^2J^F_{21}(m_d,m_s,p^2)-4J^F_{22}(m_d,m_s,p^2)
\right]\nonumber\\
&&
-\frac{3}{2\pi^2}(2m_d^2-2m_d m_s+2m_s^2-p^2)\nonumber\\
&&
\quad\quad\times
\left[m_d^2 J^F_0(m_d,m_d,p^2)+p^2 J^F_{11}(m_d,m_d,p^2)+p^2 J^F_{21}(m_d,m_d,p^2)+4J^F_{22}(m_d,m_d,p^2)\right]
\nonumber\\
&&\left.\right\}.
  \label{eqQiME}
\end{eqnarray}
The second and third terms in the brace will exist only with the dimension regularization,
and they will vanish if using other regularization methods such as proper-time regularization.
$J^F_l(m,M,p^2)$ are defined by the following integrals
\begin{eqnarray}
  && i \int \frac{d^4 q}{(2\pi)^4}
    \frac{ \{1, q^\alpha, q^\alpha q^\beta \} }{[q^2-m^2+i\epsilon][(q-p)^2-M^2+i\epsilon]}
   \nonumber\\&=&
   \{J^F_0, -p^\alpha J^F_{11},
     p^\alpha p^\beta J^F_{21}+g^{\alpha\beta}\frac{4}{D}J^F_{22} \}(m,M,p^2).
  \label{eqIntDef}
\end{eqnarray}
In the dimensional regularization, $J^F_l(m,M,p^2)$ can be expressed as
\begin{eqnarray}
  J^F_0(m,M,p^2)&=&\frac{1}{16\pi^2}[2L+1+J^i_1(m,M,p^2)],
	\label{eqJF0Def}
	\\
	J^F_{11}(m,M,p^2)&=&\frac12[\frac{J^i_0(m)}{p^2}-\frac{J^i_0(M)}{p^2}-\frac{m^2-M^2+p^2}{p^2}J^F_0(m,M,p^2)],
  \label{eqJF11Def}
	\\
	J^F_{21}(m,M,p^2)&=&\frac{1}{p^2}[J^i_0(M)+m^2 J^F_0(m,M,p^2)-4J^F_{22}(m,M,p^2)],
  \label{eqJF21Def}
	\\
	J^F_{22}(m,M,p^2)&=&\frac{D}{4(D-1)}[J^i_0(M)+m^2 J^F_0(m,M,p^2)+\frac{m^2-M^2+p^2}{2}J^F_{11}(m,M,p^2)
      -\frac12 J^i_0(M)],
  \label{eqJF22Def}
\end{eqnarray}
and the definitions of the helping functions $J^i_l$ are
\begin{eqnarray}
  J^i_0(m)&=&\frac{m^2}{16\pi^2}(2 L+\log\frac{m^2}{\mu^2}),\\
  J^i_1(m,M,p^2)&=&-2+\ln|\frac{p^2}{\mu^2}|+X_+\ln|a+X_+^2|-X_-\ln|a+X_-^2|
     \\ \nonumber &&
  +\left\{
           \begin{array}{ll}
 \displaystyle   2\sqrt{-a}(\arctanh\frac{X_+}{\sqrt{-a}}-\arctanh\frac{X_-}{\sqrt{-a}})
                            &~  p^2<0\\
 \displaystyle  \sqrt{|a|}\ln Y
                            &~  0<p^2\leq (m-M)^2\\
 \displaystyle  2\sqrt{a}(\arctan \frac{X_+}{\sqrt a}-\arctan\frac{X_-}{\sqrt a})
                            &~  (m-M)^2<p^2\leq (m+M)^2\\
  \displaystyle  \sqrt{|a|}\ln Y -2 i \pi \sqrt{|a|}
                            &~  p^2> (m+M)^2\\
           \end{array}
    \right. ,
\end{eqnarray}
where
\begin{eqnarray}
  a&=&\frac{m^2}{p^2}-\frac14(\frac{M^2}{p^2}-\frac{m^2}{p^2}-1)^2, \nonumber\\
  X_\pm&=&\frac12(\frac{M^2}{p^2}-\frac{m^2}{p^2}\pm 1), \nonumber\\
  Y&=&\left|\frac{1+X_+/\sqrt{|a|}}{1-X_+/\sqrt{|a|}}\frac{1-X_-/\sqrt{|a|}}{1+X_-/\sqrt{|a|}}\right|,\nonumber\\
  L&=&-\frac12+\left[\frac{1}{D-4}+\frac12(\gamma_E-\ln4\pi)\right].
\end{eqnarray}
\end{widetext}

\end{document}